\documentclass[conference]{IEEEtran}
\pdfoutput=1
\newcommand{\mytitle}[0]{From NoSQL Accumulo to NewSQL Graphulo:\\ Design and Utility of Graph Algorithms\\ inside a BigTable Database}
\usepackage[  
  bookmarks=false,
  hyperfootnotes=false,
  hyperindex=false,
  hidelinks,  
  pdftitle={From NoSQL Accumulo to NewSQL Graphulo: Design and Utility of Graph Algorithms inside a BigTable Database},  
  pdfauthor={Dylan Hutchison, Jeremy Kepner, Vijay Gadepally, Bill Howe}]{hyperref}
  
\usepackage{cite}
\usepackage{url}
\usepackage{tikz}
\usepackage{footnote}
\usepackage{balance}

\usepackage[ampersand]{easylist}

\graphicspath{{.}{./img/}}
\usepackage{epstopdf}
\DeclareGraphicsExtensions{.eps,.pdf,.png}
\usepackage[cmex10]{amsmath}
\usepackage{todonotes}

\usepackage{dblfloatfix} 

\usepackage{graphicx}
\usepackage{subcaption}
\usepackage{amsfonts} 
\usepackage{amssymb} 
\usepackage{authblk}

\newcommand{\matr}[1]{\ensuremath{\mathbf{#1}}} 
\newcommand{\tr}[0]{{\intercal}} 

\usepackage{mathtools}

\usepackage{siunitx}
\usepackage{tabulary}

\usepackage{multirow}
\usepackage{adjustbox}
\usepackage{array}
\newcolumntype{R}[2]{%
    >{\adjustbox{angle=#1,lap=\width-(#2)}\bgroup}%
    l%
    <{\egroup}%
}

\makeatletter
\newcommand{\removelatexerror}{\let\@latex@error\@gobble}
\makeatother

\newlength{\algspace}
\newcommand{\matlab}{\textsc{Matlab}}



\usepackage{listings}

\usepackage[linesnumbered,lined]{algorithm2e}

\usepackage{color} 
\definecolor{mygreen}{RGB}{28,172,0} 
\definecolor{mylilas}{RGB}{170,55,241}

\newcommand{\gbfont}[1]{\textsl{#1}}
\newcommand{\BuildMatrix}{\gbfont{BuildMatrix}}
\newcommand{\ExtracTuples}{\gbfont{ExtracTuples}}
\newcommand{\Transpose}{\gbfont{Transpose}}
\newcommand{\MxM}{\gbfont{MxM}}
\newcommand{\Extract}{\gbfont{Extract}}
\newcommand{\Assign}{\gbfont{Assign}}     
\newcommand{\EwiseAdd}{\gbfont{EwiseAdd}}       
\newcommand{\EwiseMult}{\gbfont{EwiseMult}}      
\newcommand{\Apply}{\gbfont{Apply}}  
\newcommand{\Reduce}{\gbfont{Reduce}}

\newcommand{\rowmode}{\textsc{row}}
\newcommand{\ewisemode}{\textsc{ewise}}

\newcommand{\jaccard}{\ensuremath{\operatorname{Jaccard}}} 
\newcommand{\ktruss}{\ensuremath{\operatorname{kTruss}}} 
\newcommand{\ttruss}{\ensuremath{\operatorname{3Truss}}} 
\newcommand{\nnz}{\ensuremath{\operatorname{nnz}}}

\begin{document}

\title{\mytitle{}}


\author[D. Hutchison et al.]
      {Dylan Hutchison$^{{\dagger}}\;$ Jeremy Kepner$^{{\ddagger}{\S}{\diamond}}\;$ 
       Vijay Gadepally$^{{\ddagger}{\S}}\;$ Bill Howe$^{{\dagger}}\;$\\
        \\
        $^{\dagger}$University of Washington$\;$
        $^{\ddagger}$MIT Lincoln Laboratory \\
        $^{\S}$MIT Computer Science \& AI Laboratory$\;$
        $^{\diamond}$MIT Mathematics Department
        \vspace{-0.4em}
      }


%

\maketitle

{\let\thefootnote\relax\footnote{\hspace{-\parindent}%
Dylan Hutchison is the corresponding
    author, reachable at dhutchis@uw.edu.
}}
\setcounter{footnote}{0}
\begin{abstract}
Google BigTable's scale-out design for distributed key-value storage 
inspired a generation of NoSQL databases. 
Recently the NewSQL paradigm emerged in response to 
analytic workloads that demand distributed computation local to data storage. 
Many such analytics take the form of graph algorithms, 
a trend that motivated the GraphBLAS initiative 
to standardize a set of matrix math kernels 
for building graph algorithms.   
In this article we show how it is possible to implement the GraphBLAS kernels
in a BigTable database
by presenting the design of Graphulo, a library for executing graph algorithms inside the Apache Accumulo database.
%
We detail the Graphulo implementation of two graph algorithms
and conduct experiments comparing their performance
to two main-memory matrix math systems. 
Our results shed insight into the conditions 
that determine when executing a graph algorithm
is faster inside a database versus an external system---in short,
that memory requirements and relative I/O are critical factors.
\end{abstract}

\IEEEpeerreviewmaketitle


\section{Introduction}
\label{sIntro}

The history of data storage and compute is long intertwined.
SQL databases facilitate
a spectrum of streaming computation inside the database server
in the form of relational queries \cite{codd1970relational}.
Server-side selection, join, and aggregation 
enable statisticians to compute correlations and run hypothesis tests 
on larger datasets \cite{ghosh1986statistical}.
The high I/O cost of geospatial queries, financial transactions,
and other more complicated computation 
motivated the concept of stored procedures for executing custom computation
in databases as early as Sybase in the 1980s \cite{stonebraker2005goes}.

The NoSQL movement marks a shift away from in-database computation,
relaxing some of the guarantees and services provided by SQL databases 
in order to provide a flexible schema and greater read/write performance
as demanded by new applications such as website indexing
\cite{grolinger2013data}. 
The Google BigTable NoSQL design in particular 
forsook relational operators in order to allow arbitrary row and column names, allow uninterpreted values,
and provide a clear model for data partitioning and layout,
all at high performance that scales with a cluster of commodity machines \cite{chang2008bigtable}. 
Several databases follow BigTable's design, including 
Apache Accumulo, Apache HBase, and Hypertable.

BigTable follows a pull-based model of computation 
called the \emph{iterator stack}:
a collection of arbitrary classes through which key-value entries flow,
beginning at a table's data sources (in-memory maps and files in a backing store)
and flowing through the logic of each iterator class 
before sending entries to a client (or a file, in the case of a compaction) after the last iterator.
The iterator stack was designed for relatively lightweight computation
such as filtering outdated values and summing values with the same keys.
As such, many applications use BigTable systems purely for storage and retrieval,
drawing on separate systems to perform computation. 

The NewSQL movement marks another shift, 
this time back toward SQL guarantees and services 
that promise efficient in-database analytics 
while retaining the flexibility and raw read/write performance of NoSQL databases \cite{grolinger2013data}.
Engineers have several reasons to consider computing inside databases rather than in external systems:
\begin{enumerate}
\item To increase data locality by co-locating computation and data storage.
The savings in data communication significantly improves performance 
for some computations.
\item To promote infrastructure reuse, 
avoiding the overhead of configuring an additional system 
by using one system for both storage and computation. 
Organizations may have administrative reasons for preferring solutions that use known, already integrated systems
rather than solutions that introduce unfamiliar or untested new systems.
\item To take advantage of database features
such as fast selective access to subsets of data along indexed dimensions.
In the case of distributed databases, algorithms running inside a database
obtain distributed execution for free, in return for cooperating with the database's access path.
\end{enumerate}

Is it possible to implement NewSQL-style analytics within a BigTable database?
A positive answer could have broad implications for organizations
that use a BigTable database to store and retrieve data
yet use a separate system for analyzing that data.
The extent of such implications depends on the relative performance 
of running analytics inside a BigTable database
versus an external system.

In this work we examine analytics that take the form of graph algorithms.
Analysts commonly interpret data as graphs
due to easy conceptualization (entities as nodes, relationships as edges), 
visualization, and applicability of graph theory (e.g. centrality and clusters) for gaining data insight.

One way to represent graphs is by its adjacency or incidence matrix.
Matrices are an excellent choice for a data structure
because they provide properties and guarantees derived from linear algebra
such as identities, commutativity, and annihilators.
These properties facilitate reasoning the correctness of algorithms and optimizations 
before any code is written, which can save considerable developer time \cite{boykin2014summingbird}.

The GraphBLAS specification is a set of signatures for matrix math kernels 
that are building blocks for composing graph algorithms \cite{kepner2015graphs}. 
These kernels provide the benefits of computing with matrices
while remaining amenable to optimization.
In particular, we show in Section~\ref{sDesign} how a critical class of optimizations, 
kernel fusion, remains possible in the GraphBLAS abstraction.

We focus our work on Graphulo, 
an effort to realize the GraphBLAS primitives inside the Apache Accumulo database. 
In past Graphulo work 
we sketched several graph algorithms in terms of the GraphBLAS kernels \cite{gadepally2015gabb}
and detailed the Graphulo implementation of Sparse Generalized Matrix Multiply \cite{hutchison2015graphulo}.
Our new contributions are:
\begin{itemize}
\item An extension of Graphulo's model for Accumulo server-side computation
to all the GraphBLAS kernels in Section~\ref{sDesign},
showing that it is possible to implement general graph analytics 
inside a BigTable system.
\item The Graphulo implementation of algorithms
to compute the Jaccard coefficients (Section~\ref{sJaccardDesign})
and the $k$-truss decomposition (Section~\ref{sKTrussAdjDesign})
of an adjacency matrix.
\item Insight into the conditions that determine when it profitable
to execute graph algorithms inside a BigTable system,
as learned from Section~\ref{sPerf}'s performance evaluation of the Jaccard and $k$-truss algorithms 
inside Accumulo and in two external main-memory matrix math systems.
\end{itemize}


Our results corroborate an emerging theme: ``no one size fits all'' \cite{stonebraker2005one};
no single system is best for every algorithm.
Our evidence shows that executing an algorithm inside a BigTable database achieves better performance 
when data does not fit in memory or the I/O cost, in terms of entries read and written, is within an order of magnitude.
Algorithms that are iterative or otherwise read and write significantly more entries in an in-database implementation
run faster in a main-memory system.

Writing an algorithm in terms of the GraphBLAS 
frees developers to execute on any GraphBLAS-enabled system.
We offer the conditions presented here 
to aid developers in choosing the best system for execution.



\section{GraphBLAS in Graphulo}
\label{sDesign}

\begin{figure}[t]
\centering
\includegraphics[width=\linewidth]{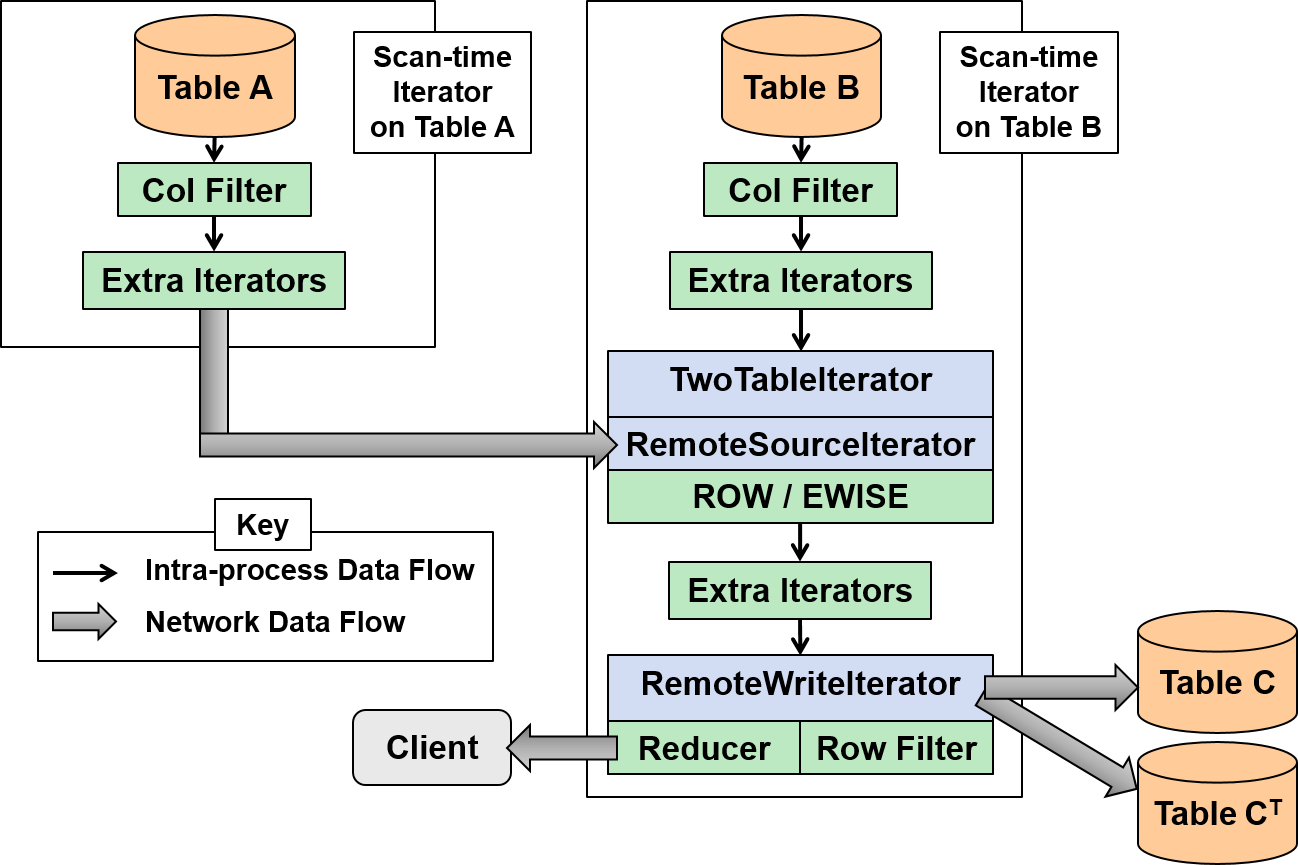}
\caption{Graphulo TwoTable template iterator stack. 
Every tablet server hosting tablets of \matr{A} and \matr{B} run this stack in parallel.\\Each GraphBLAS kernel is a special case of TwoTable.}
\label{fTwoTableOverview}
\end{figure}

Figure~\ref{fTwoTableOverview} depicts a template of the iterator stack Graphulo uses for server-side processing.
Graphulo users customize the template by calling a \texttt{TwoTable} function in the Graphulo library.
Because the TwoTable call accepts a large number of arguments in order to customize the iterator stack to 
any desired processing, Graphulo provides a number of simpler, more specialized functions
such as \texttt{TableMult} for computing an \MxM{}, \texttt{SpEWiseSum} for computing an \EwiseAdd{}, and \texttt{OneTable} for computations that take a single input table.

Once a client configures the iterator stack by calling one of Graphulo's functions, 
Accumulo instantiates copies of the stack across the nodes in its cluster.
This behavior follows BigTable's design for horizontal distribution:
\emph{tables} are divided into \emph{tablets}, each of which is hosted by a \emph{tablet server}.
Tablet servers execute a copy of the iterator stack 
that reads from each tablet of input tables \matr{A} and \matr{B} the tablet server hosts.
The same distribution applies to output tables \matr{C} and $\matr{C}^\tr$;
writes to $\matr C$ and $\matr C^\tr$ ingest into each tablet server hosting their tablets.
The whole scheme is embarrassingly parallel.
However, it is important for developers to design the rows of their table schemas intelligently
in order to evenly partition entries among their tablets 
and achieve good load balancing.

\begin{table}[t]
\centering
\begin{tabular}{ll}
GraphBLAS Kernel & Graphulo Implementation                              \\
\hline
BuildMatrix{} $(\oplus)$         & Accumulo BatchWriter                               \\
ExtracTuples{}       & Accumulo BatchScanner                              \\
MxM{} $(\oplus, \otimes)$ 
& TwoTableIterator \rowmode{} mode, performing $\matr{A}^\tr \matr{B}$            \\
EwiseMult{} $(\otimes)$          & TwoTableIterator \ewisemode{} mode                                \\
EwiseAdd{}  $(\oplus)$           & Similar to EwiseMult{}, with non-matching entries$\!\!$    \\
Extract{}            & Row and column filtering                           \\
Apply{}     $(f)$         & Extra Iterators              \\
Assign{}             & Apply with a key-transforming function            \\
Reduce{}    $(\oplus)$         & Reducer module on RemoteWriteIterator        \\           
Transpose{}          & Transpose option on RemoteWriteIterator                         \\
\end{tabular}
\caption{GraphBLAS implementation overview.}
\label{tGraphBLAStoGraphulo}
\vspace{-1.75em}
\end{table}

The boxes of Figure~\ref{fTwoTableOverview} in blue are the core components of Graphulo's iterator stack.
The RemoteSourceIterator and RemoteWriteIterator extend the Accumulo client Scanner and BatchWriter, respectively, 
to read from and write to Accumulo tables inside an iterator stack.
The TwoTableIterator merges two iterator stacks along a prefix of their keys,
depending on whether it is used for a row-wise operation like matrix multiply (\rowmode{} mode)
or an element-wise operation (\ewisemode{} mode).

Table~\ref{tGraphBLAStoGraphulo} summarizes the GraphBLAS kernels
and how parts of the TwoTable stack realize them.
We discuss each kernel separately to show
how they are individually implemented, but in practice
kernel fusion is critical for performance.

To fuse a set of kernels is to execute them in one step, without 
writing the intermediary matrices between them.
BLAS packages have a long history of 
fused kernels, stemming from the original GEMM call
which fuses matrix multiplication, matrix addition, scalar multiplication,
and transposition \cite{dongarra1990set}.

Kernel fusion is particularly important in the case of BigTable databases
because writing out intermediary tables implies extra round trips to disk.
It is possible to fuse kernels in Graphulo until a sort is required,
since sorting is a blocking operation.
Related research efforts are exploring additional GraphBLAS optimizations,
such as decomposing kernels into finer-grained tasks
that better utilize  parallel architecture \cite{wolf2015task}.

The symbols $\oplus$, $\otimes$, and $f$ in Table~\ref{tGraphBLAStoGraphulo}
indicate user-defined functions supplied as parameters to the GraphBLAS kernels.
In Graphulo these functions take the form of user-provided Java classes. 
Graphulo provides interfaces that make writing these operations easy
provided they abide by a contract following the structure of a semi-ring:
$0 \otimes a = 0$, $0 \oplus a = a$, $f(0) = 0$,
associativity, and idempotence and distributivity in certain cases.
Developers are free to insert general iterators or break these contracts,
so long as they understand their role in Accumulo's distributed execution and lazy summing.

The following sections discuss 
Graphulo's representation of matrices,
Graphulo's TwoTable iterators, 
and how those iterators implement the GraphBLAS kernels.

\subsection{Matrices in Accumulo}
We represent a matrix as a table in Accumulo.
The row and column qualifier portions of an Accumulo key 
store the row and column indices of a matrix entry.
Values are uninterpreted bytes that can hold any type, not necessarily numeric.

Other portions of the Accumulo key---column family, visibility, and timestamp---are 
available for applications to use.
Many applications use the visibility portion of a key for cell-level security,
restricting operations to run only on those entries a user has permission to see.
Applications can define how visibilities propagate through the Graphulo iterators.

The column family provides a flexible grouping capability.
For example, it is possible to store two matrices inside the same Accumulo table 
by distinguishing them with separate column families.
This could be useful when the two matrices are frequently accessed together, 
since Accumulo would store them in the same tablets with alignment on rows.
Applications may also leverage Accumulo's locality groups in order to store certain 
column families separately, which is similar to how Accumulo separately stores distinct tables
but on a fine-grained, columnar level.

Zero-valued entries need not be stored in the Accumulo table
in order to efficiently store sparse matrices.
We treat the concept of ``null'' the same as the concept of ``zero''
for both storage and processing. 
However, it is not forbidden to store zero entries;
a zero entry may spuriously arise during processing 
when 3, -5, and 2 are summed together, for example,
though Graphulo prunes such entries as an optimization by default.
We encourage developers not to rely on the presence or absence of zero entries
for algorithm correctness.



\subsection{BuildMatrix{} and ExtracTuples{}}
The \BuildMatrix{} and \ExtracTuples{} GraphBLAS functions
are the constructors and destructors of matrices from or to a set of ``triples'':
a list of (row, column, value) entries.

Accumulo already supports these operations by means of its BatchWriter and BatchScanner
client interfaces. Constructing a matrix is as simple as writing each tuple to an Accumulo table.
Destructing a matrix is as simple as scanning a table.

If the list of triples for \BuildMatrix{} contains 
multiple values for the same row and column,
users commonly define a function $\oplus$
to store the sum of colliding values in the matrix.
Accumulo combiner iterators achieve this behavior 
by lazily summing duplicate values according to $\oplus$ during compactions and scans.\footnote{BigTable 
systems do not run iterators on entries immediately as they are ingested in order to maximize write performance.}
Default behavior ignores all but one of them.



\subsection{MxM{}}
Graphulo matrix multiplication was previously explained in \cite{hutchison2015graphulo}.
As a brief summary, Graphulo computes the matrix multiplication $\matr{A} \matr{B}$
by placing a RemoteSourceIterator, TwoTableIterator, and RemoteWriteIterator
on a BatchScan of table $\matr{B}$.
The RemoteSourceIterator scans the transpose table $\matr{A}^\tr$.
The TwoTableIterator aligns entries from $\matr{A}^\tr$ and $\matr{B}$ 
along the row portion of their keys, iterating both inputs in lockstep until a matching row is found.
A user-defined Java class implementing a $\otimes$ function
multiplies entries from matching rows according to the outer product algorithm.
Partial products flow into the RemoteWriteIterator, which sends entries to the result table 
via an Accumulo BatchWriter.
A user-defined iterator class implementing a $\oplus$ function
lazily sums values on the result table during its next scan or compaction.

We call TwoTableIterator's workflow during \MxM{} its \rowmode{} mode.
More advanced uses of \rowmode{} mode may provide 
a user-defined strategy for processing two aligned rows of data.
The \jaccard{} implementation described in Section \ref{sJaccardDesign}, for example,
uses custom row processing to fuse the \EwiseAdd{} of three \MxM{} kernels
by performing additional multiplications.

\subsection{EwiseAdd{} and EwiseMult{}}
The element-wise GraphBLAS kernels are similar to \MxM{},
except that TwoTableIterator runs in \ewisemode{} mode
and table $\matr A$ is not treated as its transpose.
During \ewisemode{} mode, TwoTableIterator aligns entries from $\matr A$ and $\matr B$
along their row, column family, and column qualifier.
The \EwiseMult{} kernel passes matching entries to a user-defined $\otimes$ function.

The \EwiseAdd{} kernel acts on both matching and non-matching entries.
Non-matching entries pass directly to the RemoteWriteIterator.
Matching entries pass to a ``multiplication function'' that implements the $\oplus$ addition logic.

\subsection{Extract{}}
The \Extract{} kernel stores the subset of a matrix into a new matrix.
Ranges of row and column indices specify the subset.

Graphulo implements \Extract{} via row and column filtering on the ranges of indices. 
The client passes these ranges to Accumulo's tablet servers
by transmitting serialized options attached to iterator configuration data.
The RemoteWriteIterator handles row filtering by seeking the iterators
above it to only read entries from the indexed rows.
Column filtering occurs at an iterator right after reading entries from the table.

The difference between row and column filtering is due to Accumulo's design as a row-store database.
Whereas filtering rows is efficient since all data is accessed and stored by row,
column filtering requires reading entries from all columns and discarding those
outside the column indices.
Future work could optimize column filtering for locality groups when present.

\subsection{Apply{} and Assign{}}
The \Apply{} kernel applies a function $f$ to every entry of a matrix.
The kernel assumes that $f(0) = 0$ so that processing need only run on nonzero entries.
Applying $f$ in Accumulo takes the form of an extra iterator implementing $f$ 
that can be placed at any point in an iterator stack.
It is easy to fuse \Apply{} with other kernels
by including the iterator for $f$ at appropriate locations.

Graphulo supports both stateless and stateful \Apply{} functions $f$,
with the caveat that stateful functions must cope with distributed execution 
wherein multiple instances of $f$ run concurrently, each one seeing a portion of all entries. 

The \Assign{} kernel assigns a matrix to a subset of another matrix 
according to a set of row and column indices.
An \Apply{} iterator implements \Assign{} by transforming the keys of entries to their corresponding keys in the new matrix.

\subsection{Reduce{}}
The \Reduce{} kernel gathers information of reduced dimensionality about a matrix
via a process similar to user-defined aggregation \cite{cohen2006user}.
Often this information is a scalar, such as the set of unique values that occur in a matrix 
or the number of entries whose value exceeds a threshold.
These scalars may be used inside control structures, 
such as how \ktruss{} in Section \ref{sKTrussAdjDesign}
multiplies matrices until the number of partial products at each multiplication does not change, indicating convergence.

Many \Reduce{} use cases transmit data to the client for control purposes 
rather than write data to a new table.
In order to accommodate these cases and also facilitate fusing \Reduce{} into other kernels, 
we integrated \Reduce{} into the RemoteWriteIterator
by coupling it with a ``Reducer object'' 
that processes every entry the RemoteWriteIterator sees.

The Reducer object implements the signature of a commutative monoid.
It has a ``zero'' state that is the state of the Reducer before processing any entries.
The reducer ``sums in'' entries one at a time via user-defined $\oplus$ logic.
Once the RemoteWriteIterator finishes processing entries for the tablet it is running on,
it asks the Reducer for the result of reducing all the entries the Reducer has seen,
and it forwards that result to the client 
through the standard BatchScanner channel on which the client instantiated the whole TwoTable stack.
The client combines local results from reducing each tablet to obtain the global result of \Reduce{}.

The above scheme works because a \Reduce{} call's result is typically small. 
It is feasible to fit the result into tablet server memory and transmit it to the client.
If the result is not small, it may be wiser to store results at the server in a new table.
\Reduce{} calls that store results at the server, such as summing the columns of a matrix 
into a vector, can be implemented via \Apply{} and a $\oplus$ iterator on the result table,
or sometimes as an \MxM{} with a constant matrix.

\subsection{Transpose{}}
The \Transpose{} kernel switches the row and column of entries from a matrix.
While \Transpose{} could be implemented as an \Apply{}, we found that it is used so frequently
that it is worth making it a built-in option of the RemoteWriteIterator.

\section{Algorithms}
\label{sAlgorithms}

\subsection{Vertex Similarity: Jaccard Coefficients}
\label{sJaccardDesign}
Vertex similarity is an important metric for applications such as link prediction \cite{liben2007link}. 
One measure for similarity between two vertices is their Jaccard coefficient. 
This quantity measures the overlap of two vertices' neighborhoods in an unweighted, undirected graph. 
For vertices $v_{i}$ and $v_{j}$ where $N(v)$ denotes the neighbors of vertex $v$, the Jaccard coefficient is defined as
\[
\matr J_{ij} =\frac{\mid N(v_{i}) \bigcap N(v_{j}) \mid}{\mid N(v_{i}) \bigcup N(v_{j}) \mid}
\]

Gadepally et al formulated an algorithm in terms of the GraphBLAS to compute the Jaccard coefficients 
between all vertices for a graph represented as an unweighted, undirected adjacency matrix \cite{gadepally2015gabb}. 
Algorithm~\ref{algJaccard} summarizes their formulation.
They employ an optimization that restricts computation to the upper triangle of the adjacency matrix,
taking advantage of the symmetry in the Jaccard coefficient definition.
The notation $\operatorname{triu}(\cdot,1)$ borrows \matlab{} syntax for 
taking the strict upper triangle of a matrix.

Line 1 stores the degrees of each vertex in a vector $\matr{d}$.
For the Graphulo implementation, we assume these degrees are pre-computed 
inside a separate degree table in Accumulo.
Computing degree tables is often performed during data ingest,
since fast access to degree information is useful for query planning, 
load balancing, filtering, and other analytics \cite{kepner2013d4m}.

\begin{figure}[t]
\vspace{-0.5em}
\centering
\includegraphics[width=0.88\linewidth]{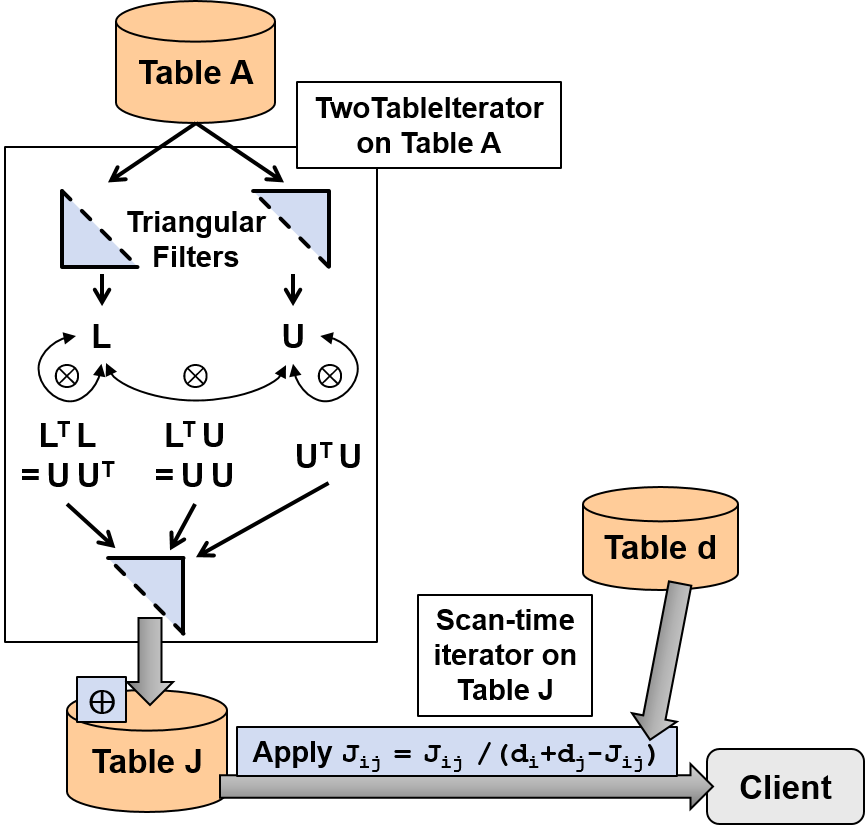}
\caption{Visualization of the \jaccard{} algorithm. The three `$\otimes$' symbols highlight the fused \MxM{} kernel below them.}
\label{fJaccardViz}
\vspace{-0.5em}
\end{figure}

Graphulo computes the Jaccard coefficient matrix $\matr J$ in one fused \MxM{} call
that Figure~\ref{fJaccardViz} highlights.
The inputs to the \MxM{} are the lower and upper triangles $\matr L$ and $\matr U$ of $\matr A$,
which we obtain by applying strict lower and upper triangle filters.

Given inputs $\matr L$ and $\matr U$, the TwoTableIterator computes\footnote{Recall that Graphulo transposes the left input to \MxM{} calls. 
The transpose of lower triangular matrix $\matr{L}$ is the upper triangular matrix $\matr{U}$.}
$\matr{L}^\tr \matr{U} = \matr{U} \matr{U}$.
We could compute the other required products $\matr{U} \matr{U}^\tr$ and $\matr{U}^\tr \matr{U}$ in a similar fashion with two separate multiplication calls.
However, all the information we need to compute $\matr{U} \matr{U}^\tr$ and $\matr{U}^\tr \matr{U}$ is available during the computation of $\matr{U} \matr{U}$,
which suggests an opportunity to fuse the three matrix multiplications together.

\begin{algorithm}[t]
	\DontPrintSemicolon
	\SetCommentSty{textit}
	\KwIn{Unweighted, undirected adjacency matrix $\matr A$}
	\KwOut{Upper triangle of Jaccard coefficients $\matr J$}
	\begingroup
	\def\J{\matr{J}}
	\def\d{\matr{d}}
	\def\U{\matr{U}}
	\def\A{\matr{A}}
	$\d = \operatorname{sum}(\A)$ \tcp*{pre-computed in degree table} 
	$\U = \operatorname{triu}(\A, 1)$ \tcp*{strict upper triangle filter}
	$\J = \operatorname{triu}(\U\U + \U\U^\tr + \U^\tr\U, 1)$\tcp*{fused \MxM{}}
	\ForEach{nonzero entry $\J_{ij} \in \J$}{
		$\J_{ij}=\J_{ij}/(\d_{i} + \d_{j} - \J_{ij})$\tcp*{stateful \Apply{} on \matr{J}}
	}
	\endgroup
\vspace{5pt}
\caption{\jaccard{}\newline Comments describe the Graphulo implementation.}
\label{algJaccard}
\end{algorithm}

We implement the fusion of $\matr{U} \matr{U}^\tr + \matr{U}^\tr \matr{U} + \matr{U} \matr{U}$ by providing a custom row multiplication function to TwoTableIterator
that computes all three products at once. 
In addition to multiplying pairs of entries in the Cartesian product of matching rows between $\matr{L}$ and $\matr{U}$ to compute $\matr{L}^\tr \matr{U} = \matr{U} \matr{U}$,
the custom function multiplies pairs of entries in the Cartesian product of its left input $\matr{L}$ with itself to compute $\matr{L}^\tr \matr{L} = \matr{U} \matr{U}^\tr$,
as well as its right input $\matr{U}$ with itself to compute $\matr{U}^\tr \matr{U}$.
The additional multiplications also run on rows of $\matr{L}$ and $\matr{U}$ that do not match, as in the pattern of an \EwiseAdd{} kernel.
After passing the entries from all three matrix multiplications 
through a further strict upper triangle filter (line 3),
the RemoteWriteIterator sends them to the result table $\matr J$
with an iterator that sums colliding values.

After the \MxM{} operation completes, indicating that all partial products from the three matrix multiplications have written to table $\matr J$,
we add a stateful \Apply{} iterator to the scan scope of $\matr J$ that computes lines 4--6.
The iterator performs a kind of broadcast join, scanning the degree table \matr{d} into the memory
of tablet servers hosting \matr{J}'s tablets in order to efficiently compute line 5.
Holding $\matr d$ in memory is feasible because it is significantly smaller than the original table $\matr A$.


As a final optimization to \MxM{} that applies when multiplying a matrix with itself,
we used the ``deep copy'' feature of Accumulo iterators
to duplicate the iterator stack on $\matr{A}$ 
and use it for both inputs to TwoTableIterator.
This optimization eliminates the need for a RemoteSourceIterator,
which saves entry serialization and coordination with Accumulo's master.
While it did not significantly increase performance 
during Section \ref{sPerf}'s single-node evaluation,
we anticipate the deep copy technique will have larger impact in a multi-node setting.

\subsection{Truss Decomposition}
\label{sKTrussAdjDesign}
The $k$-truss of an unweighted, undirected simple graph 
is the largest subgraph in which every edge is part of at least $k-2$ triangles.
Computing the $k$-truss is useful for focusing large graph analysis onto a cohesive subgraph
that has a hierarchical structure one can vary with $k$ \cite{wang2012truss}.

Gadepally et al formulated a GraphBLAS algorithm to compute the $k$-Truss 
of a graph represented by an unweighted, undirected incidence matrix \cite{gadepally2015gabb}.
The algorithm iteratively deletes edges that are part of fewer than $k-2$ triangles
until all edges are part of at least $k-2$ triangles.
Iteration is a particularly challenging feature for an Accumulo implementation 
since Accumulo stores intermediary tables on disk.

We adapted the 
algorithm to run on a graph's adjacency matrix 
in Algorithm~\ref{algkTrussAdjacencyOpt}.
The \nnz{} call in line 9 computes the number of nonzero entries in a matrix.
The product $\matr A \matr A$ in line 5 computes the number of triangles each edge is part of.

The `\%' symbol in line 6 indicates remainder after integer division, and so line 6 deletes even values.
Because $\matr{A}$'s nonzero values are odd (specifically, the value 1)
and because partial products from $2\matr A \matr A$ are even (specifically, the value 2), 
line 6 effectively filters out entries from $\matr{B}$ that are not present in $\matr{A}$.
Line 7 deletes edges that are part of fewer than $k-2$ triangles
after ``undoing'' the $+1$ from $\matr A$ and the $\times 2$ from $2\matr A \matr A$.
The $|\matr{B}|_0$ in line 8 indicates the zero norm of $\matr{B}$, which sets nonzero values to 1.

\begin{algorithm}[tb]
	\DontPrintSemicolon
	\SetCommentSty{textit}
	\KwIn{Unweighted, undirected adjacency matrix $\matr A_0$, 
	integer $k$}
	\KwOut{Adjacency matrix of $k$-truss subgraph $\matr A$}
	\begingroup
	\def\B{\matr{B}}
	\def\A{\matr{A}}
	$z' = \infty, \matr{A} = \matr{A}_0$\tcp*{table clone}
	\Repeat{$z == z'$\tcp*[f]{client controls iteration}}{
		$z = z'$\;
		$\B = \A$\tcp*{table clone}
		$\B = \B + 2\A\A$\tcp*{\MxM{} with $a \otimes b = 2$ if $a, b \neq 0$}
		$\B(\B \,\%\, 2 == 0) = 0$\tcp*{filter on \matr{B}}
		$\B((\B - 1)/2 < k-2) = 0$\tcp*{filter on \matr{B}}
		$\A = |\B|_0$\tcp*{\Apply{} on \matr{B}; switch $\matr{A} \leftrightarrow \matr{B}$}
		$z' = \operatorname{nnz}(\A)$\tcp*{\Reduce{}, gathering \nnz{} at client}
	}
 	\endgroup
\vspace{5pt}
\caption{\ktruss{}\newline Comments describe the Graphulo implementation.} 
\label{algkTrussAdjacencyOpt}
\end{algorithm}

We specifically designed Algorithm~\ref{algkTrussAdjacencyOpt} to minimize the number of intermediary tables
in order to minimize round-trips to disk.
The insight that allows us to reduce the number of intermediary table writes,
from two in a naive formulation to one as presented here,
is finding a way to distinguish edges that are part of at least $k-2$ triangles 
but not present in the original graph from edges present in the original graph.
A naive method to distinguish  edges 
is by computing the \EwiseMult{} $\matr A \otimes \matr B$
where $\otimes$ is ``logical and''.
We eliminate the \EwiseMult{} by playing tricks with parity in lines 5--7 as described above.


The Graphulo \ktruss{} implementation uses two temporary tables 
$\matr A$ and $\matr B$.
We initialize $\matr A$ by cloning input table $\matr A_0$.\footnote{Cloning Accumulo tables 
is a cheap operation because no data is copied; 
Accumulo simply marks the cloned table's Hadoop RFiles as shared with the clonee after flushing entries in memory to disk.}

Each iteration begins with cloning $\matr A$ into $\matr B$ as in line 4.
Implementing line 4 with a table clone is an optimization that avoids rewriting the entries in $\matr A$.
Graphulo constructs an \MxM{} iterator stack to multiply $\matr A$ with itself\footnote{Recall that 
$\matr A^\tr = \matr A$ because $\matr A$ is the adjacency matrix for an undirected graph. 
The product $\matr A^\tr \matr A$ is the same as $\matr A \matr A$.} and sum the result into $\matr B$ as in line 5.
The \MxM{} stack includes a $\otimes$ function that evaluates to 2 on nonzero inputs,
as well as an extra iterator following $\otimes$ that filters out entries along the diagonal as another optimization,
which is correct since the $k$-truss is defined on simple graphs.
A standard $\oplus$ iterator on $\matr B$ sums partial products.

After the \MxM{} stack completes, indicating that $\matr B$ contains every partial product,
Graphulo places an additional iterator on $\matr B$ after the $\oplus$ iterator
to filter out entries that are even (line 6) or fail the $k$-Truss condition (line 7).
A final iterator on $\matr B$ sets nonzero values to 1 as in line 8.

A \Reduce{} call computes $\nnz(\matr{A})$ in line 9 by counting entries.
The algorithm has converged and may terminate per line 10
once $\operatorname{nnz}(\matr A)$ does not change between iterations.

The temporary tables switch roles between iterations, 
deleting the old $\matr A$ before switching $\matr A$ with $\matr B$ and again cloning $\matr A$ into $\matr B$.
After the last iteration, we rename the new $\matr A$ to the designated output table via a clone and delete.




\section{Performance}
\label{sPerf}

In this section we conduct an experiment to 
(1) provide a single-node evaluation of Graphulo on the \jaccard{} and \ttruss{} (fixing $k = 3$ in \ktruss{}) algorithms and
(2) gain insight into when it is profitable to execute the algorithms inside Accumulo 
versus an external system.

We compare Graphulo to \emph{main-memory} external systems
because they are among the best options for computing on a subgraph
(e.g. cued analytics \cite{Reuther09cloudcomputing}),
a use case databases accelerate by creating table indexes.
Such subgraphs lie on the threshold of fitting into memory wherein main-memory computation is feasible;
we therefore choose problem sizes that also lie on the threshold of fitting in memory. 
We consider both sparse and dense matrix systems
because dense systems generally perform orders of magnitude faster but have severe memory constraints,
whereas sparse systems offer intermediate performance in exchange for handling larger matrices.

Specifically, we choose two open-source main-memory matrix math systems for comparison:
the dynamic distributed dimensional data model (D4M) \matlab{} library for sparse matrices \cite{kepner2012dynamic}
and the Matrix Toolkits Java (MTJ) library for dense matrices \cite{MTJ}.
D4M provides an associative array API 
that maps to databases like Accumulo 
and calls \matlab{}'s sparse matrix functions \cite{kepner2015associative};
D4M has frequently been used to analyze subsets of database tables that fit in memory.
We choose D4M for the role of sparse matrix math
because it is one of the best GraphBLAS alternatives to Accumulo for single-node computation; 
GraphMat \cite{sundaram2015graphmat} (single-node) 
as well as CombBLAS \cite{bulucc2011combinatorial} and GraphPad \cite{anderson2016graphpad} (distributed)
are also candidates.
MTJ is one of many Java libraries that provide high-performance matrix math \cite{arndt2009towards}.
We choose MTJ for the role of dense matrix math for its off-the-shelf ease of use. 

Following our past experimental setup~\cite{hutchison2015graphulo},
we evaluate the performance of \jaccard{} and \ttruss{} by measuring runtime
while varying two parameters: problem size and computational resources.
Varying these parameters allows us to gauge 
both weak scaling (varying problem size while holding computational resources constant)
and strong scaling (varying computational resources while holding problem size constant).

We control problem size via the size of input matrices,
and we control computational resources via the number of tablets on input and output tables.
The number of tablets controls how many threads Accumulo uses for reading, writing, and iterator processing.
In our single-node setup, this relates to how well Accumulo uses the 8 cores it has available.
We limit the number of tablets to 1 or 2, since Accumulo runs enough threads to fully use all 8 cores at 2 tablets.

We set the number of tables on input tables by compacting them prior to each experiment.
For 2 tablet experiments, we choose a split point that evenly divides each input table
and pre-split output tables on the same splits.\footnote{In the case of Graphulo \ktruss{}, all intermediary tables use the same splits as the input table since cloning a table preserves its splits.}
In order to prevent Accumulo from creating additional splits during 
larger SCALE experiments (17 for \jaccard{}, 16 for \ttruss{}),
we increased the \texttt{table.split.threshold} parameter to 5 GB.

Most users deploy Accumulo on a sizable cluster.
Graphs as large as 70 trillion edges (1.1 PB) have been analyzed 
by clusters as large as 1200 machines and 57.6 TB collective memory \cite{burkhardt2015cloud}.
This work evaluates single-node performance as a proof of concept  
and an indicator for how multi-node performance might scale.

\begin{figure*}[t]
\centering
\begin{minipage}[c]{1.0\columnwidth}
\includegraphics[width=\linewidth]{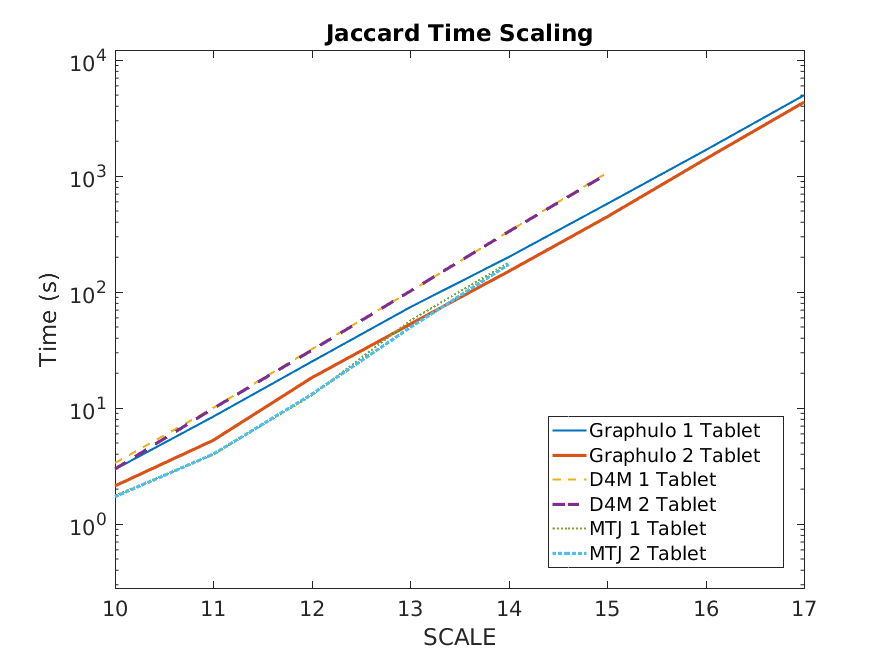}
\caption{\jaccard{} experiment processing time.}
\label{fJaccardTime}
\end{minipage}
\hfil
\begin{minipage}[c]{1.0\columnwidth}
\includegraphics[width=\linewidth]{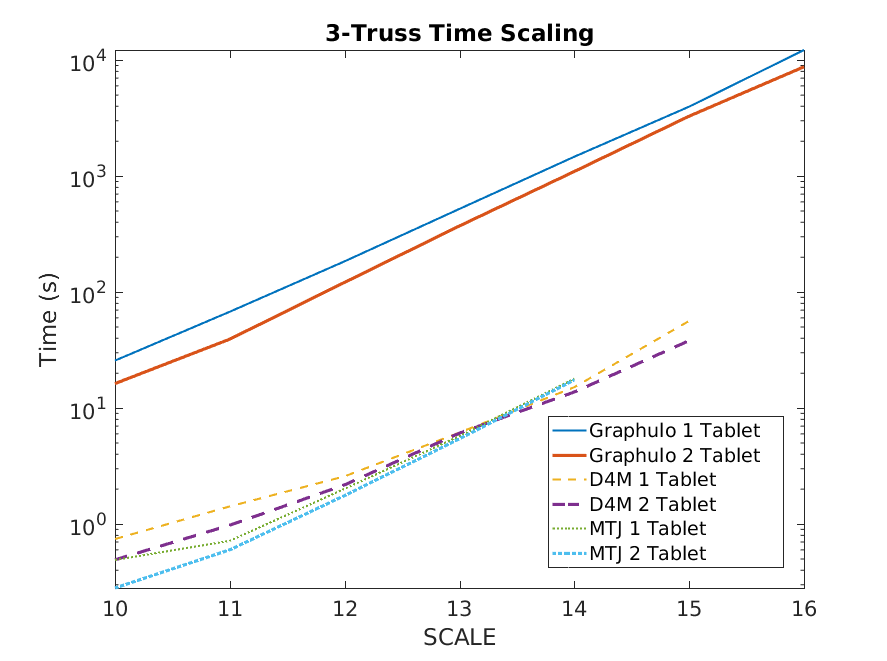}
\caption{\ttruss{} experiment processing time.}
\label{fKTrussAdjTime}
\end{minipage}
\end{figure*}

\let\stimes\times
\renewcommand{\times}[0]{{\,\stimes{}\,}}

\begin{table*}[t]
\centering
\begin{minipage}[l]{1.0\linewidth}
\centering
\begin{tabular}{c|c|c|c|r|c@{$\quad\;$}c|c@{$\quad$}c|c@{$\quad$}c}
\adjustbox{angle=30,lap={1.75em}{\width-3.8em},raise=-0.75em,scale=0.95}{SCALE} & $\operatorname{nnz}(\matr{A})$ & \begin{tabular}{@{}c@{}}$\operatorname{nnz}($ \\ $\!\!\jaccard(\matr{A}))\!\!\!$ \end{tabular} & \begin{tabular}{@{}c@{}} Partial \\ Products \end{tabular} & \begin{tabular}{@{}c@{}} Graphulo \\ Overhead \end{tabular} & \begin{tabular}{@{}c@{}}Graphulo \\1 Tablet \end{tabular} & \begin{tabular}{@{}c@{}}Graphulo \\2 Tablets \end{tabular} & \begin{tabular}{@{}c@{}}D4M \\1 Tablet \end{tabular} & \begin{tabular}{@{}c@{}}D4M \\2 Tablets \end{tabular}   & \begin{tabular}{@{}c@{}}MTJ \\ 1 Tablet \end{tabular}  & \begin{tabular}{@{}c@{}}MTJ \\ 2 Tablets \end{tabular} \\
\hline
10 & \num{21038.000} & \num{214762.000} & \num{1011708.000} & 4.7x & \num{2.969} & \num{2.135} & \num{3.364} & \num{2.988} & \num{1.762} & \num{1.722} \\
\hline
11 & \num{45240.000} & \num{706522.000} & \num{3103514.000} & 4.4x & \num{8.461} & \num{5.287} & \num{10.070} & \num{9.960} & \num{3.994} & \num{4.014} \\
\hline
12 & \num{96716.000} & \num{2179554.000} & \num{9293469.000} & 4.3x & \num{25.171} & \num{18.294} & \num{32.243} & \num{31.562} & \num{12.886} & \num{13.173} \\
\hline
13 & \num{204056.000} & \num{6749240.000} & \num{27148963.000} & 4.0x & \num{74.183} & \num{52.960} & \num{101.521} & \num{101.983} & \num{56.829} & \num{49.577} \\
\hline
14 & \num{426152.000} & \num{20209303.000} & \num{77718934.000} & 3.8x & \num{200.603} & \num{151.074} & \num{333.851} & \num{332.575} & \num{179.179} & \num{173.285} \\
\hline
15 & \num{883418.000} & \num{60724113.000} & \num{221956191.000} & 3.7x & \num{576.980} & \num{445.879} & \num{1067.778} & \num{1046.129} &  &  \\
\hline
16 & \num{1818976.000} & \num{176702345.000} & \num{619875458.000} & 3.5x & \num{1675.095} & \num{1405.334} &  &  &  &  \\
\hline
17 & \num{3729062.000} & \num{515569673.000} & \num{1724038721.000} & 3.3x & \num{4986.014} & \num{4339.107} &  &  &  &  \\
\end{tabular}
\caption{\jaccard{} experiment statistics. Graphulo is competitive and better scales due to low overhead.}
\label{tJaccardTable}
\end{minipage}
\vspace{1em}

\begin{minipage}[l]{1.0\linewidth}
\centering
\begin{tabular}{c|c|c|c|r|c@{$\quad\;$}c|c@{$\quad$}c|c@{$\quad$}c}
\adjustbox{angle=30,lap={1.75em}{\width-3.8em},raise=-0.75em,scale=0.95}{SCALE} & $\operatorname{nnz}(\matr{A})$ & \begin{tabular}{@{}c@{}}$\operatorname{nnz}($ \\ $\!\ttruss(\matr{A}))\!$ \end{tabular} & \begin{tabular}{@{}c@{}} Partial \\ Products \end{tabular} & \begin{tabular}{@{}c@{}} Graphulo \\ Overhead \end{tabular} & \begin{tabular}{@{}c@{}}Graphulo \\1 Tablet \end{tabular} & \begin{tabular}{@{}c@{}}Graphulo  \\2 Tablets \end{tabular} & \begin{tabular}{@{}c@{}}D4M \\1 Tablet \end{tabular} & \begin{tabular}{@{}c@{}}D4M \\2 Tablets \end{tabular}   & \begin{tabular}{@{}c@{}}MTJ \\ 1 Tablet \end{tabular}  & \begin{tabular}{@{}c@{}}MTJ \\ 2 Tablets \end{tabular} \\
\hline
10 & \num{21038.000} & \num{20250.000} & \num{5940204.000} & 293.3x & \num{25.692} & \num{16.274} & 0.74 & 0.49 & 0.49 & 0.28 \\
\hline
11 & \num{45240.000} & \num{43488.000} & \num{12208688.000} & 280.7x & \num{67.791} & \num{39.316} & \num{1.421} & 0.98 & 0.72 & 0.60 \\
\hline
12 & \num{96716.000} & \num{92020.000} & \num{54540918.000} & 592.7x & \num{184.403} & \num{121.446} & \num{2.581} & \num{2.179} & \num{2.017} & \num{1.763} \\
\hline
13 & \num{204056.000} & \num{192860.000} & \num{159197278.000} & 825.5x & \num{522.009} & \num{372.184} & \num{6.156} & \num{6.090} & \num{5.741} & \num{5.437} \\
\hline
14 & \num{426152.000} & \num{399132.000} & \num{455259412.000} & 1140.6x & \num{1470.584} & \num{1097.113} & \num{15.154} & \num{13.773} & \num{17.925} & \num{17.493} \\
\hline
15 & \num{883418.000} & \num{819656.000} & \num{1297106806.000} & 1582.5x & \num{3973.623} & \num{3289.679} & \num{56.477} & \num{38.162} &  &  \\
\hline
16 & \num{1818976.000} & \num{1669822.000} & $3.62 \times 10^9$ & 2167.0x & \num{12233.052} & \num{8767.558} &  &  &  &  \\
\end{tabular}
\caption{\ttruss{} experiment statistics. D4M and MTJ execute faster, assuming sufficient memory, due to high overhead.}
\label{tKTrussAdjTable}
\end{minipage}

Graphulo overhead is defined as how many times more entries Graphulo writes into Accumulo than D4M or MTJ. Runtimes are listed in seconds.
\vspace{-0.0em}
\end{table*}

\renewcommand{\times}[0]{\stimes}

We conduct the experiments on a Linux Mint 17.2 laptop with 16 GB RAM, two dual-core Intel i7 processors,
and a 512 GB SSD. 
Atop single-instance Hadoop 2.4.1 and ZooKeeper 3.4.6,
we allocated 2 GB of memory to an Accumulo tablet server with room to grow to 3 GB,
1 GB for native in-memory maps and 256 MB for data and index cache.
We ran a snapshot build of Accumulo 1.8.0 at commit \texttt{9aac9c0}
in order to incorporate a bugfix that affects Graphulo's stability \cite{ACCUMULO-4229}. 

We use the Graph500 unpermuted power law graph generator \cite{bader2006designing} 
to create random input adjacency matrices
whose first rows are high-degree ``super-nodes''
and whose subsequent rows exponentially decrease in degree.
Power law distributions are widely used to model real world graphs \cite{gadepally2015using}.

The generator takes SCALE and EdgesPerVertex parameters, creating matrices with 2\textsuperscript{SCALE} 
rows and EdgesPerVertex $\times$ 2\textsuperscript{SCALE} entries.
We fix EdgesPerVertex to 16 and use SCALE to vary problem size. 
In order to create undirected and unweighted adjacency matrices without self-loops,
we merge the generated matrix with its transpose, ignore duplicate entries, and filter out the diagonal.
These modifications slightly change the input graphs' edge count;
see the \nnz(\matr A) column in Tables \ref{tJaccardTable} and \ref{tKTrussAdjTable} for exact counts.

Figures~\ref{fJaccardTime} and \ref{fKTrussAdjTime} plot the runtime for 
Graphulo, D4M, and MTJ on 1 and 2 tablets and various problem sizes. 
D4M's sparse matrices and MTJ's dense matrices
exceed our machine's memory at SCALE 16 and 15, respectively.

Figure~\ref{fJaccardTime} indicates competitive Graphulo performance on the \jaccard{} algorithm.
Graphulo always outperforms D4M and runs on par with MTJ 
for problem sizes that D4M and MTJ can hold in memory.
Figure~\ref{fKTrussAdjTime} shows an order of magnitude faster D4M and MTJ performance on the \ttruss{} algorithm.

In order to analyze the disparity in performance between \jaccard{} and \ttruss{}, 
we present additional experimental information 
in Tables \ref{tJaccardTable} and \ref{tKTrussAdjTable}.
The ``\nnz(\jaccard{}(\matr A))'' and ``\nnz(\ttruss{}(\matr{A}))'' columns
list the number of nonzero entries in the result from their algorithms.
D4M and MTJ write exactly this many entries into Accumulo
because they compute the complete result in memory and insert it as is.

Recall that a partial product is a multiplied value $a*b$ computed during a matrix multiplication.
The ``Partial Products'' column in Tables \ref{tJaccardTable} and \ref{tKTrussAdjTable}
lists the total number of partial products computed during each algorithm
(not including entries filtered out by \jaccard{}'s strict upper triangle filter and \ktruss{}'s no-diagonal filter).
Graphulo writes exactly this many entries into Accumulo
because it implements the outer product matrix multiply algorithm;
Graphulo writes individual partial products to Accumulo and defers summing them to a $\oplus$ iterator
that runs during compactions and scans.
In contrast, D4M and MTJ pre-sum partial products before writing to Accumulo in a manner similar to the inner product algorithm.
Unfortunately an inner product formulation that pre-sums partial products 
is infeasible for Graphulo 
because, unlike a main-memory system, Graphulo does not assume it can hold a whole table in memory 
and would therefore need to re-read an input table many times, including many re-reads from disk.

We define the \emph{Graphulo overhead} as how many times more entries Graphulo writes into Accumulo than D4M or MTJ.
For reference, the Graphulo overhead of multiplying two power law matrices is 2.5--3x, decreasing with matrix size \cite{hutchison2015graphulo}.
Table~\ref{tJaccardTable} shows that \jaccard{} has a similar Graphulo overhead of 3--5x, also decreasing with matrix size.
At this low an overhead, it makes sense that Graphulo outperforms D4M and MTJ
because the performance gain from computing inside Accumulo outweighs the cost of writing additional entries.

Table~\ref{tKTrussAdjTable} shows a drastically different Graphulo overhead for \ttruss{} of 280--2200x, increasing with matrix size.
The source of the additional overhead is that Graphulo writes out an intermediary table at each iteration,
including all partial products of $\matr A \matr A$ before applying the filter in lines 6 and 7 of Algorithm~\ref{algkTrussAdjacencyOpt}.
D4M and MTJ do not need to write intermediary values since they hold them in memory.
At this high an overhead, D4M and MTJ are better places to execute \ttruss{}.

\begin{figure}[t]
\centering
\includegraphics[width=\linewidth]{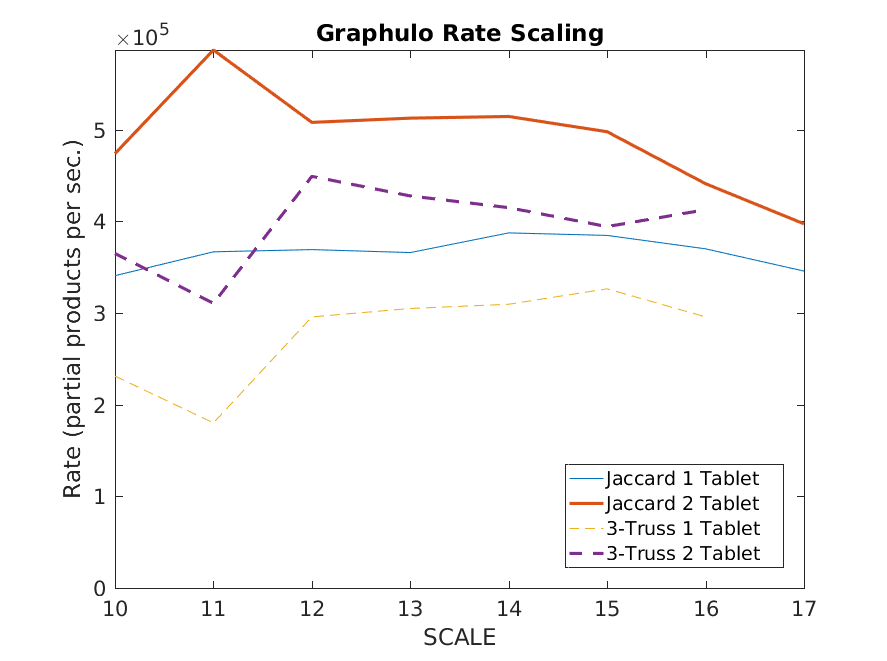}
\caption{Graphulo Processing Rate during \jaccard{}, \ttruss{}.}
\label{fGraphuloRate}
\vspace{-1em}
\end{figure}

In terms of weak and strong scaling, Graphulo actually performs quite well if we account for its overhead.
Figure~\ref{fGraphuloRate} plots Graphulo's processing rate during \jaccard{} and \ttruss{}.
We express processing rate as the number of partial products written to Accumulo divided by algorithm runtime
so that we can compare processing rate to Accumulo insert rates.

For reference, the largest cited peak single-node write rates into Accumulo 
are on the order of 400k per second \cite{kepner2014achieving}.
We believe the reason why the \jaccard{} 2-tablet rate exceeds the single-node record
is due to its fused \MxM{} call: writing partial products from 3 multiplications
while reading a single table.

Both algorithms show relatively horizontal lines, 
indicating constant processing rate as problem size varies.
Two-tablet rates are about 1.5 times greater than one-tablet rates,
indicating a potential for strong scaling 
that a multi-node experiment could further evaluate in future work.

\section{Related Work}
\label{sRelated}


One way to characterize the difference between \jaccard{} and \ktruss{}'s performance inside Accumulo
is as the difference between data- and state-parallel algorithms.
A \emph{data-parallel} operation easily runs on data partitioned among several threads or machines.  
A \emph{state-parallel} operation runs more efficiently on data gathered in one location, 
either because of data correlation, the need for global calculation, 
or the presence of iteration which induces communication between every round.  
 
Many distributed databases, Accumulo included, efficiently run data-parallel workloads 
but are poorly suited to state-parallel workloads.
The solution proposed in Section 4.3 of the UDA-GIST framework 
\cite{li2015uda}---to gather the entire state into a single node during the iterative portion of the workflow---is essentially the solution that D4M and MTJ implement.

The Graphulo \Reduce{} implementation is similar in design to the GLADE aggregation framework 
for implementing generalized linear aggregates (GLAs) inside parallel databases \cite{rusu2012glade}.
Both follow a pattern of accumulating information inside a thread-local ``reducer object''
and coalescing the objects from multiple threads and nodes by serializing and merging them,
in order to gather a fully reduced result at one location.

Multi-platform query optimizers such as Rheem \cite{agrawal2016road} and Musketeer \cite{gog2015musketeer} evaluate equivalent implementations of an algorithm on different data processing environments, in order to select the best one relative to a cost model for execution.
It seems clear that a multi-platform optimizer could evaluate and compare
two plans for executing a graph algorithm,
one using Graphulo and one using a main-memory system.
Less clear is whether the overhead conditions described in Section \ref{sPerf},
under which execution on one system outperforms the other,
could be encoded into a cost model.
Should such a cost model exist, we could spare developers 
the burden of analyzing algorithms 
in order to determine the system that executes them best.
The BigDAWG polystore framework takes a different approach, 
introducing `scope' clauses for developers to manually specify the system 
for executing portions of an algorithm 
and `cast' clauses for moving data between system \cite{duggan2015bigdawg}.

Cheung et al applied a program synthesis technique called \emph{verified lifting}
in order to recognize recognize fragments of legacy code
which operate on data obtained from a database query
and could be pushed into the database in order to achieve better performance
\cite{cheung2013optimizing}.
Similar to multi-platform query optimizers,
verified lifting facilitates automatic rewriting 
of algorithm fragments to execute on the best-suited system,
assuming an accurate cost model.
Their difference is that optimizers take higher-level queries as input, 
usually in a form that can be parsed into a logical algebra,
whereas verified lifting abstracts a high-level specification
from general-purpose code. 

PipeGen is an initiative to reduce the cost of data transfer between systems 
by generating efficient binary transfer protocols 
and injecting them into systems' import and export facilities \cite{haynes2016pipegen}.
Should efforts like PipeGen gain momentum,
computing in specialized external systems 
may gain feasibility for algorithms with otherwise prohibitive data transfer cost.

\section{Conclusion}
\vspace{-0.5em}
In this work we detail how Graphulo's design enables 
executing the GraphBLAS kernels inside the Accumulo database.
We cover the implementation of two graph algorithms
and show how to optimize them for in-database execution via kernel fusion.
Experiments comparing their performance to two main-memory matrix math systems
show that I/O, in terms of database reads and writes,
is a critical factor for determining whether an algorithm executes best 
inside a database or in an external system, assuming enough available memory.

In future work we aim to extend our analysis to a multi-node setting
and include additional systems such as Spark and CombBLAS.
Characterizing the traits of algorithms that determine their performance on different data processing systems
is an exciting first step toward robust cross-system algorithm optimization,
using each system where it performs best.

\vspace{-0.75em}
\section*{Acknowledgment}
\vspace{-0.5em}
The authors thank Brandon Myers and Zachary Tatlock for their energetic feedback,
Timothy Weale for his engineering camaraderie,
the whole Graphulo team bridging Seattle and Cambridge,
and Alan Edelman, Sterling Foster, Paul Burkhardt, and Victor Roytburd.
This material is based on work supported by the National Science Foundation
Graduate Research Fellowship Program under Grant No. DGE-1256082.\hspace{-0.3em}
\vspace{-0.69em}





%

%
%

\bibliographystyle{IEEEtran}

\bibliography{10_bibliography}

\balance

\end{document}